\documentclass[times]{smrauth}

\usepackage{moreverb}
\usepackage[colorlinks,bookmarksopen,bookmarksnumbered,citecolor=red,urlcolor=red]{hyperref}
\usepackage{amsmath}
\usepackage{bbm}
\usepackage{amssymb}
\usepackage{graphicx}
\usepackage{listings}
\usepackage{multirow}
\usepackage{algorithmic}
\usepackage{hyperref}
\usepackage{url}
\usepackage{verbatim}
\usepackage{courier}
\usepackage{caption}
\usepackage{colortbl}
\usepackage{color}
\usepackage{xcolor}
\usepackage{booktabs}
\usepackage{setspace}
\usepackage{booktabs}

\newcommand{\word}[1]{\ensuremath{\mathop{\,\mathsf{#1}\,}}}
\newcommand{\type}[1]{\ensuremath{\text{#1}}}

\newcommand{\figref}{Fig. \ref}

\newtheorem{definition}{Definition}

\newcommand{\model}[0]{\ensuremath{\mathcal{M}}}

\newcommand{\trace}[0]{\ensuremath{\omega = (s_0,t_0)(s_1,t_1)\cdots(s_{N-1},t_{N-1})\cdots}}

\newcommand{\prefixtrace}[1]{\ensuremath{\omega_{#1} = (s_0,t_0)(s_1,t_1)\cdots(s_{#1-1},t_{#1-1})}}

\newcommand{\suffixtrace}[1]{\ensuremath{\omega^{#1} = (s_{#1},t_{#1})(s_{#1+1},t_{#1+1})\cdots}}

\newcommand{\temporal}[0]{\ensuremath{\mathcal{T}_r}}

\newcommand\BibTeX{{\rmfamily B\kern-.05em \textsc{i\kern-.025em b}\kern-.08em
T\kern-.1667em\lower.7ex\hbox{E}\kern-.125emX}}

\def\volumeyear{2016}

\begin{document}

\runningheads{Ngo and Legay}{Statistical Model Checking for Probabilistic SystemC Models}

\title{Formal Verification of Probabilistic SystemC Models with 
Statistical Model Checking}

\author{Van Chan Ngo\affil{1}\corrauth Axel Legay\affil{2}}

\address{\affilnum{1}Computer Science Department, Carnegie Mellon University, Pittsburgh, PA 15213, USA \break
\affilnum{2}Inria Rennes - Bretagne Atlantique, Rennes, 35042, France}

\corraddr{Computer Science Department, Carnegie Mellon University, Pittsburgh, PA 15213, USA.\\
E-mail: \textit{channgo@cmu.edu}}

\begin{abstract}
Transaction-level modeling with SystemC has been very successful in describing the behavior of embedded systems by providing high-level executable models, in which many of them have inherent probabilistic behaviors, e.g., random data and unreliable components. It thus is crucial to have both quantitative and qualitative analysis of the probabilities of system properties.

Such analysis can be conducted by constructing a formal model of the system under verification and using Probabilistic Model Checking (PMC). However, this method is infeasible for large systems, due to the state space explosion. In this article, we demonstrate the successful use of Statistical Model Checking (SMC) to carry out such analysis directly from large SystemC models and allow designers to express a wide range of useful properties.

The first contribution of this work is a framework to verify properties expressed in Bounded Linear Temporal Logic (BLTL) for SystemC models with both timed and probabilistic characteristics. 

Second, the framework allows users to expose a rich set of user-code primitives as atomic propositions in BLTL. Moreover, users can define their own fine-grained time resolution rather than the boundary of clock cycles in the SystemC simulation.

The third contribution is an implementation of a statistical model checker. It contains an automatic monitor generation for producing execution traces of the model-under-verification (MUV), the mechanism for automatically instrumenting the MUV, and the interaction with statistical model checking algorithms.
\end{abstract}


\maketitle

\footnotetext[2]{Work partly done while the author was at Inria Rennes - Bretagne Atlantique, Rennes, France.}

\section{Introduction}
\label{sec:introduction}
Transaction-level modeling (TLM) with SystemC has become increasingly prominent in describing the behavior of embedded systems~\cite{cch99}, e.g., System-on-Chips (SoCs). Complex electronic components and software control units can be combined into a single model, enabling simulation of the whole system at once. In many cases, models have probabilistic and non-deterministic characteristics, e.g, random data and reliability of the system's components. Hence, it is crucial to evaluate the quantitative and qualitative analysis of the probabilities of system properties.

\subsection{Motivation}
We consider a safety-critical system, e.g., a control system for an air-traffic, automotive, or medical device. The reliability and availability of the system can be modeled as a stochastic process, in which it exhibits both timed and probabilistic characteristics. For instance, the reliability and availability model of an embedded control system~\cite{knp07} that contains an input processor connected to groups of sensors, an output processor connected to groups of actuators, and a main processor that communicates with the I/O processors through a bus. Suppose that the sensors, actuators, and processors can fail and the I/O processors have transient and permanent faults. When a transient fault occurs in a processor, rebooting the processor repairs the fault. The times to failure and the delay of reboot are exponentially distributed. Thus, the reliability of the system can be modeled by a \emph{Continuous-Time Markov Chain} (CTMC)~\cite{tks82,gaa87} (a special case of a discrete-state \emph{stochastic process} in which the probability distribution of the next state depends only on the current state). The analysis can be quantifying the probabilities or rates of all safety-related faults: How likely is it that the system is available to meet a demand for service? What is the probability that the system repairs after a failure (e.g., the system conforms to the existent and prominent standards such as the \emph{Safety Integrity Levels} (SILs))?

In order to conduct such analysis, a general approach is modeling and analyzing a probabilistic model of the system (e.g., Markov chains, stochastic processes), in which the algorithm for computing the measures in properties depends on the class of systems being considered and the logic used for specifying the property. Many algorithms with the corresponding mature tools are based on model checking techniques that compute the probability by a numerical approach~\cite{brv04,cg04,rkn04,hwz08}. Timed automata with mature verification tools such as UPPAAL~\cite{lpy05} are used to verify real-time systems. For a variety of probabilistic systems, the most popular modeling formalism is Markov chain or Markov decision processes, for which \emph{Probabilistic Model Checking} (PMC) tools such as PRISM~\cite{hkn06} and MRMC~\cite{khh09} can be used. PMC is widely used and has been successfully applied to the verification of a range of timed and probabilistic systems. One of the main challenges is the complexity of the algorithms in terms of execution time and memory space due to the size of the state space that tends to grow exponentially, also known as the \emph{state space explosion}. As a result, the analysis is infeasible. In addition, these tools cannot work directly with the SystemC source code, meaning that a formal model of SystemC model needs to be provided.

An alternative way to evaluate these systems is \emph{Statistical Model Checking} (SMC), a simulation-based approach. Simulation-based approaches produce an approximation of the value to be evaluated, based on a finite set of system's executions. Clearly, compared to the numerical approaches, a simulation-based solution does not provide an exact answer. However, users can tune the statistical parameters such as the absolute error and the level of confidence, according to the requirements. Simulation-based approaches do not construct all the reachable states of the model-under-verification (MUV), thus they require far less execution time and memory space than the numerical approaches. For some real-life systems, they are the only option~\cite{ykn06} and have shown the advantages over other methods such as PMC~\cite{hwz08,jcj09}.

\subsection{Contribution}
In this article, we demonstrate the successful use of SMC to carry out directly qualitative and quantitative analysis for probabilistic temporal properties of large SystemC models. SMC also allows designers to express a wide range of useful properties. The work makes the following contributions.
\begin{itemize}
\item We propose a framework to verify bounded temporal properties for SystemC models with both timed and probabilistic characteristics. The framework contains two main components: a \emph{monitor} observing a set of execution traces of the MUV and a \emph{statistical model checker} implementing a set of hypothesis testing algorithms. We use techniques proposed by Tabakov et al.~\cite{tva10} to automatically generate the monitor. The statistical model checker is implemented as a plugin of the checker Plasma Lab~\cite{bcl13}, in which the properties to be verified are expressed in \emph{Bounded Linear Temporal Logic} (BLTL).

\item We present a method that allows users to expose a rich set of user-code primitives in form of atomic propositions in BLTL. These propositions help users exposing the state of the SystemC simulation kernel and the full state of the SystemC source code model. In addition, users can define their own fine-grained time resolution that is used to reason about the semantics of the logic expressing the properties rather than the boundary of clock cycles in the SystemC simulation.

\item We implement a tool, in which the properties of interest are expressed using BLTL. The various features of the tool including
automatic generation of monitor for generating execution traces of the MUV, mechanism to instrument automatically the MUV, and the interaction with statistical model checking algorithms are presented. We demonstrate our approach through a running example, as well as the performance of the framework through some experiments.

\item To make our verification perform on all possible execution orders of the MUV, we implement a probabilistic scheduler. Specifically, the source of process scheduling is the evaluation phase in the current SystemC scheduler implementation, in which one of the runnable processes is executed. Every time at the evaluation phase, given a set of $N$ runnable processes, our probabilistic scheduler randomly chooses one of these processes to execute. The random algorithm is implemented by generating a random integer number uniformly over the range $[0,N-1]$.
\end{itemize}
\section{Background}
\label{sec:background}
This section introduces the SystemC modeling language and reviews the main features of statistical model checking for stochastic processes as well as bounded linear temporal logic which is used to express system properties.
\subsection{SystemC Language}
\subsubsection{Language Features}
SystemC\footnote{IEEE Standard 1666-2005} is a C++ library~\cite{glm02} providing primitives for modeling hardware and software systems at the level of transactions. Every SystemC model can be compiled with a standard C++ compiler to produce an executable program called executable specification. This specification is used to simulate the system behavior with the provided event-driven simulator. A SystemC model is a hierarchical composition of modules (\word{sc\_module}). Modules are building blocks of SystemC design, they are like modules in Verilog~\cite{tmo08}, classes in C++. A module consists of an interface for communicating with other modules and a set of processes running concurrently to describe the functionality of the module. An interface contains ports (\word{sc\_port}) that are similar to the hardware pins. Modules are interconnected using either primitive channels (e.g., the signals, \word{sc\_signal}) or hierarchical channels via their ports. Channels are data containers that generate events in the simulation kernel whenever the contained data changes.

Processes are not hierarchical, so no process can call another process directly. A process is either a thread or a method. A thread process (\word{sc\_thread}) can suspend its execution  by calling the library statement \word{wait} or any of its variants. When the execution is resumed, it will continue from that point. Threads run only once during the execution of the program and are not expected to terminate. On the other hand, a method process (\word{sc\_method}) cannot suspend its execution by calling \word{wait} and is expected to terminate. Thus, it only returns the control to the kernel when reaching the end of its body.

An event is an instance of the SystemC event class (\word{sc\_event}) whose occurrence triggers or resumes the execution of a process. All processes which are suspended by waiting for an event are resumed when this event occurs, we say that the event is notified. A module's process can be sensitive to a list of events. For example, a process may suspend itself and wait for a value change of a specific signal. Then, only this event occurrence can resume the execution of the process. In general, a process can wait for an event, a combination of events, or for an amount of time to be resumed.

In SystemC, integer values are used as discrete time model. The smallest quantum of time that can be represented is called \emph{time resolution}, meaning that any time value smaller than the time resolution will be rounded off. The available time resolutions are \word{femtosecond}, \word{picosecond}, \word{nanosecond}, \word{microsecond}, \word{millisecond}, and \word{second}. SystemC provides functions to set time resolution and declare a time object.

\begin{figure}[!th]
\includegraphics[width=0.65\textwidth]{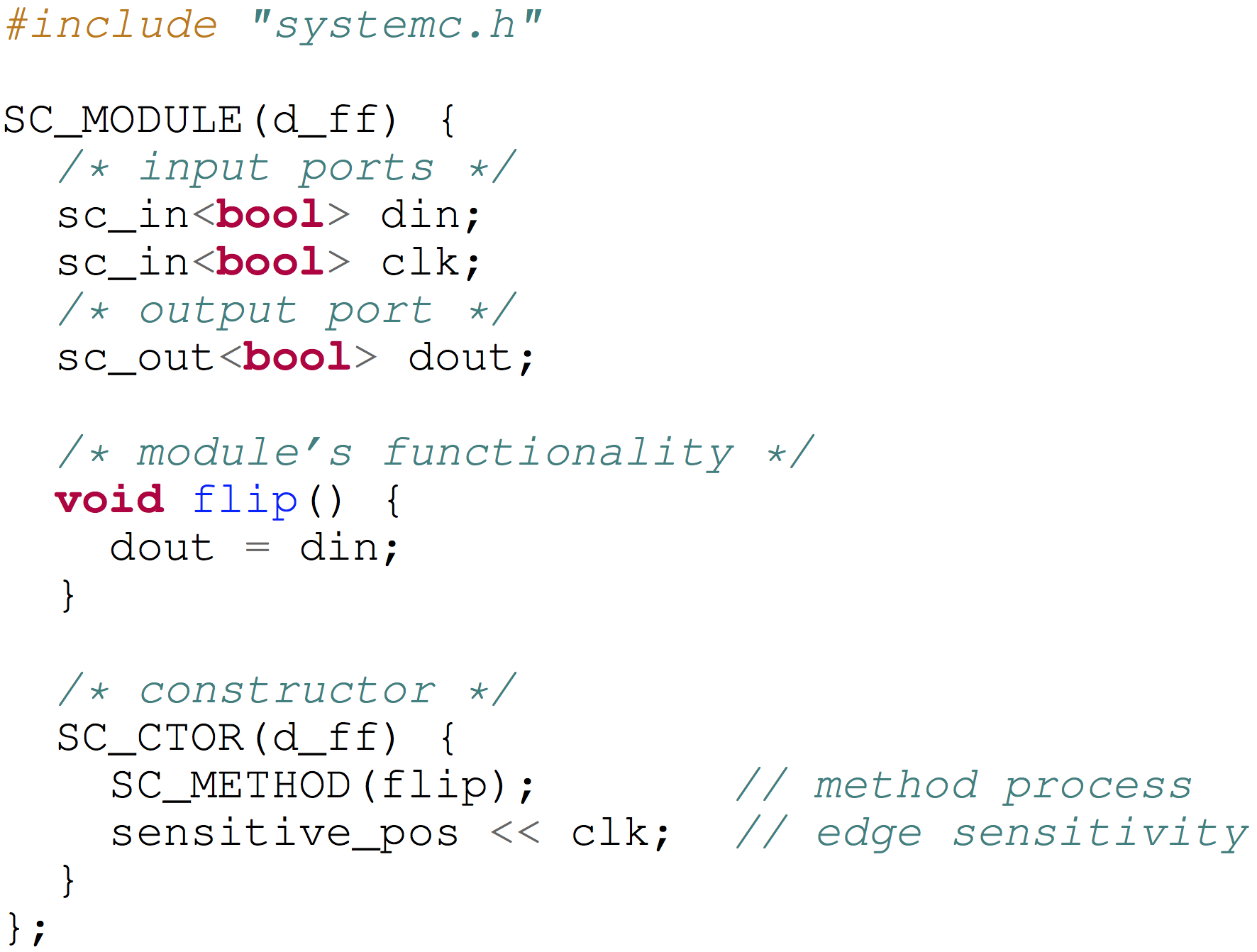}
\caption{An Implementation of Flip-Flop in SystemC.}
\label{lst:dff}
\end{figure}

The example in \figref{lst:dff} describes a simple delay flip-flop, in which \word{sc\_in} and \word{sc\_out} are predefined primitive input and output ports. The sensitivity list contains the edge sensitivity \word{sensitive\_pos} specified on \word{clk}, which indicates that only on the rising edge of \word{clk} does \word{din} get transferred to \word{dout}. The sensitivity list and process instantiation are defined in the class constructor (\word{sc\_ctor}).
\subsubsection{Simulation Kernel}
\begin{figure}[!th]
\includegraphics[width=0.87\textwidth]{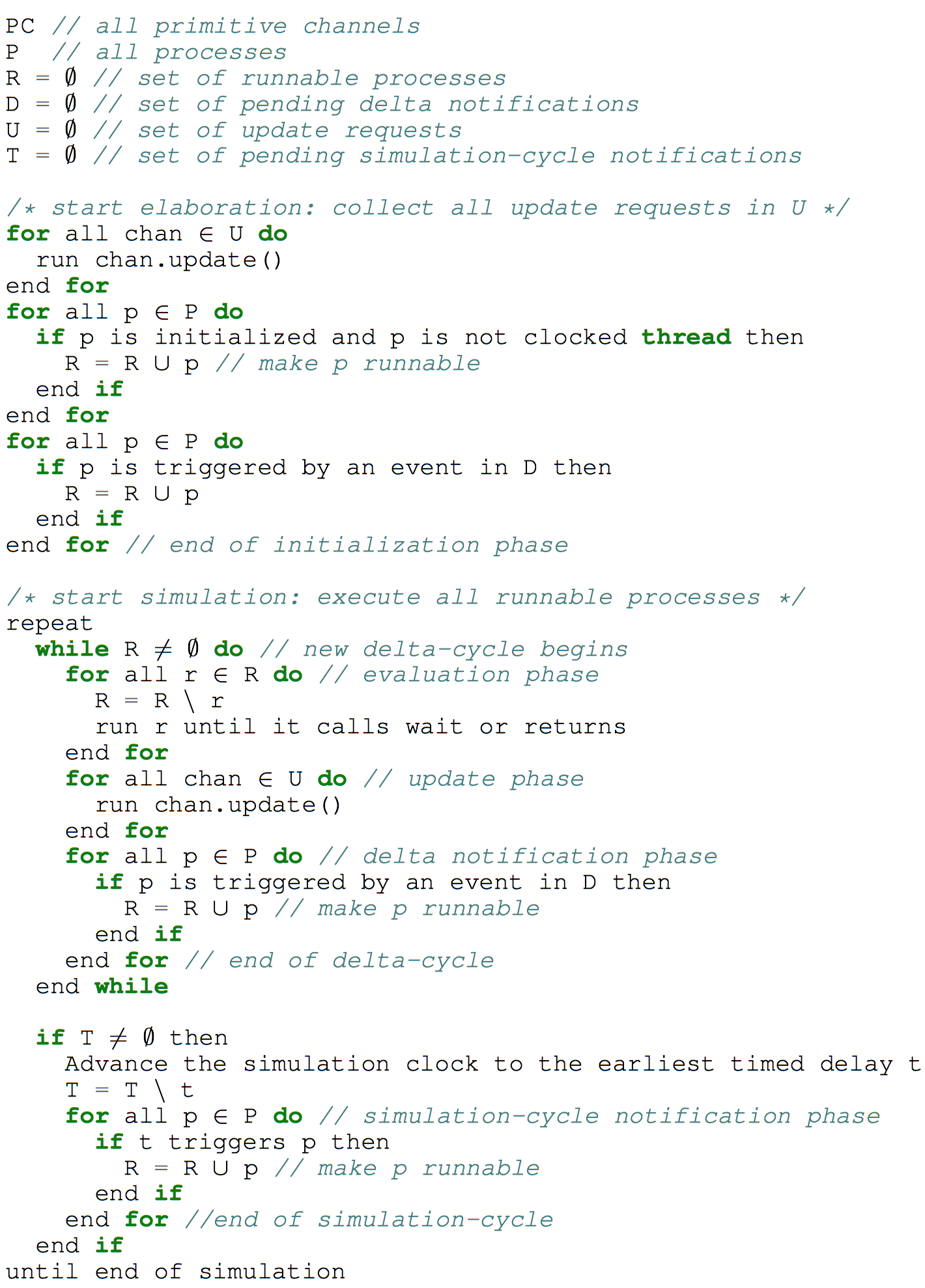}
\caption{Semantics of SystemC kernel.}
\label{lst:simulationsemantics}
\end{figure}

The SystemC simulator is an event-driven simulation~\cite{sysc,mrh01}. It establishes a hierarchical network of finite number of parallel communicating processes which are under the supervision of the distinguished simulation kernel process. Only one process is dispatched by the scheduler to run at a time point, and the scheduler is non-preemptive, that is, the running process returns control to the kernel only when it finishes executing or explicitly suspends itself by calling \word{wait}. Like hardware modeling languages, the SystemC scheduler supports the notion of delta-cycles~\cite{lsu93}. A delta-cycle lasts for an infinitesimal amount of time and is used to impose a partial order of simultaneous actions which interprets zero-delay semantics. Thus, the simulation time is not advanced when the scheduler processes a delta-cycle. During a delta-cycle, the scheduler executes actions in two phases: the \emph{evaluate} and the \emph{update} phases. The semantics of the scheduler is given in \figref{lst:simulationsemantics}. Each phase of the scheduler is explained as follows.
%
\begin{itemize}
\item \emph{Initialize}. During the initialization, each process is executed once unless it is turned off by calling \word{dont\_initialize()}, or until a synchronization point (e.g., a \word{wait}) is reached. The order in which these processes are executed is unspecified.
\item \emph{Evaluate}. The kernel starts a delta-cycle and runs all processes that are ready to run one at a time. In this same phase a process can be made ready to run by an event notification.
\item \emph{Update}. Execute any pending calls to \word{update()} resulting from calls to \word{request\_update()} in the evaluate phase. Note that a primitive channel uses \word{request\_update()} to have the kernel call its \word{update()} function after the execution of processes.
\item \emph{Delta-cycle Notification}. The kernel enters the delta notification phase where notified events trigger their dependent processes. Note that immediate notifications may make new processes runnable during step (2). If so the kernel loops back to step (2) and starts another evaluation phase and a new delta-cycle. It does not advance simulation time.
\item \emph{Simulation-cycle Notification}. If there are no more runnable processes, the kernel advances simulation time to the earliest pending timed notification. All processes sensitive to this event are triggered and the kernel loops back to step (2) and starts a new delta-cycle. This process is finished when all processes have terminated or the specified simulation time is passed.
\end{itemize}
\subsection{Bounded Linear Temporal Logic}
We here recall the syntax and semantics of BLTL~\cite{sva05}, an extension of \emph{Linear Temporal Logic} (LTL) with time bounds on temporal operators. A formula $\varphi$ is defined over a set of atomic propositions $\word{AP}$ which can be either $\word{true}$ or $\word{false}$. A BLTL formula is defined by the following BNF grammar, in which $\word{true}$ and $\word{false}$ are Boolean constants. The time bound $\word{T}$ is an amount of time or a number of states in an execution trace.
\begin{displaymath}
\varphi ::= \word{true} \mid \word{false} \mid p \in \word{AP} \mid \varphi_1 \wedge \varphi_2 \mid \neg \varphi \mid \varphi_1 \word{U_{\leq T}} \varphi_2
\end{displaymath}
The temporal modalities $\word{F}$ (the ``eventually'', sometimes in the future) and $\word{G}$ (the ``always'', from now on forever) can be derived from the ``until'' $\word{U}$ as follows.
\begin{displaymath}
\word{F_{\leq T}} \varphi = \word{true} \word{U_{\leq T}} \varphi \text{ and } \word{G_{\leq T}} \varphi = \neg \word{F_{\leq T}} \neg \varphi
\end{displaymath}
The semantics of BLTL is defined with respect to execution traces of a model $\model$. Let $\trace$, where $N \in \mathbb{N}$, be a sequence of pairs of states and time points, where the model stays in the state $s_i$ for $t_{i+1} - t_i = \delta t_i \in \mathbb{R}_{\geq 0}$ duration of time. The sequence $\omega$ is called an execution trace of $\model$. We use $\prefixtrace{k}$ and $\suffixtrace{k}$ to denote the prefix and suffix of $\omega$ respectively. We denote the fact that $\omega$ satisfies the BLTL formula $\varphi$ by $\omega \models \varphi$. The semantics is given as follows.
\begin{itemize}
\item $\omega^k \models \word{true}$ and $\omega^k \not \models \word{false}$
\item $\omega^{k} \models p, p \in \word{AP}$ iff $p \in L(s_k)$, where $L(s_k) \subseteq \word{AP}$ is the set of atomic propositions which are $\word{true}$ in state $s_k$
\item $\omega^{k} \models \varphi_1 \wedge \varphi_2$ iff $\omega^k \models \varphi_1$ and $\omega^k \models \varphi_2$
\item $\omega^k \models \neg \varphi$ iff $\omega^k \not \models \varphi$
\item $\omega^k \models \varphi_1 \word{U_{\leq T}} \varphi_2$ iff there exists $i \in \mathbb{N}$ such that $\omega^{k+i} \models \varphi_2$, $\Sigma_{0 < j \leq i}(t_{k+j} - t_{k+j-1}) \leq \word{T}$, and for each $0 \leq j < i, \omega^{k+j} \models \varphi_1$
\end{itemize}
Here is a simple temporal property expressed in BLTL that can be verified with SMC:
\begin{displaymath}
\varphi = \word{G_{\leq T_1}} (\word{A} \rightarrow \word{F_{\leq T_2}}(\word{B} \word{U_{\leq T_3}} \word{C}))
\end{displaymath}
The meaning of $\word{Pr}(\varphi)$ is: What is the probability that during the $\word{T_1}$ time units of the system operation, if $\word{A}$ holds then, starting from $\word{T_2}$ time units after, $\word{B}$ happens before $\word{C}$ within $\word{T_3}$ time units?
\subsection{Statistical Model Checking}
\label{sec:smc}
Let $\model$ be the formal model of the MUV (e.g., a stochastic process, a CTMC) and $\varphi$ be a property expressed as a BLTL formula. BLTL ensures that the satisfaction of a formula is decidable in a finite number of steps. The statistical model checking~\cite{ldb10} problem consists in answering the following questions:
\begin{itemize}
\item \emph{Qualitative Analysis}. Is the probability that $\model$ satisfies $\varphi$ greater or equal to a threshold $\theta$ with a specific level of statistical confidence?
\item \emph{Quantitative Analysis}. What is the probability that $\model$ satisfies $\varphi$ with a specific level of statistical confidence?
\end{itemize}
They are denoted by $\model \models \word{Pr}_{\geq \theta}(\varphi)$ and $\model \models \word{Pr}(\varphi)$, respectively. Many statistical model checkers have been implemented~\cite{bcl13,yhl05,hwz08,jcj09} that have shown their advantages over other methods such as PMC on several case studies.

The key idea is to associate each execution trace of $\model$ with a discrete random Bernoulli variable $B_i$. The outcome for $B_i$, denoted by $b_i$, is $1$ if the trace satisfies $\varphi$ and $0$ otherwise. The predominant statistical method for verifying $\model \models \word{Pr}_{\geq \theta}(\varphi)$ is based on \emph{hypothesis testing}. Let $p = \word{Pr}(\varphi)$, to determine whether $p \geq \theta$, we test the hypothesis $\word{H_0}: p \geq p_0 = \theta + \delta$ against the alternative hypothesis $\word{H_1}: p \leq p_1 = \theta - \delta$ based on the observations of $B_i$. The size of \emph{indifference region} is defined by $p_0 - p_1 = 2\delta$. If we take acceptance of $\word{H_0}$ to mean acceptance of $\word{Pr}_{\geq \theta}(\varphi)$ as true and acceptance of $\word{H_1}$ to mean rejection of $\word{Pr}_{\geq \theta}(\varphi)$ as false, then we can use \emph{acceptance sampling} (e.g., Younes in~\cite{you05} has proposed two solutions, called \emph{single sampling plan} and \emph{sequential probability ratio test}) to verify $\word{Pr}_{\geq \theta}(\varphi)$. An acceptance sampling test with \emph{strength} $(\alpha,\beta)$, where $\alpha, \beta \in [0, 1]$, guarantees that $\word{H_1}$ is accepted with probability at most $\alpha$ when $\word{H_0}$ holds and $\word{H_0}$ is accepted with probability at most $\beta$ when $\word{H_1}$ holds, called a Type-I error (\emph{false negative}) and Type-II error (\emph{false positive}), respectively. The probability of accepting $\word{H_0}$  is therefore at least $1 - \alpha$. We say that $1 - \alpha$ is the confidence of the hypothesis testing.

To answer the quantitative question, $\model \models \word{Pr}(\varphi)$, an alternative statistical method, based on \emph{estimation} instead of hypothesis testing, has been developed. For instance, the probability estimations are based on results derived by Chernoff and Hoeffding bounds~\cite{hoe63}. This approach uses $n$ observations $b_1,...,b_n$ to compute an approximation of $p$: $\tilde{p} = \frac{1}{n}\Sigma^{n}_{i=1}b_i$. If we require that the probability that difference between the approximation and the actual value is bounded by $\delta \in [0, 1]$ is at least $1 - \alpha$, where $\alpha \in [0, 1]$, or $\word{Pr}[|\tilde{p} - p| < \delta] \geq 1 - \alpha$. Then, based on the theorem of Hoeffding, the number of observations which is determined from the absolute error $\delta$ and the confidence $1 - \alpha$ is $n = \lceil \frac{1}{2\delta^2}log\frac{2}{\alpha} \rceil$.

Although SMC can only provide approximate results with a user-specified level of statistical confidence, it is compensated for by its better scalability and resource consumption. Since the models to be analyzed are often approximately known, an approximate result in the analysis of desired properties within specific bounds is quite acceptable. SMC has recently been applied in a wide range of research areas including software engineering~\cite{hwz08} (e.g., verification of critical embedded systems), system biology, or medical area~\cite{jcj09}.

\section{A Running Example}
\label{sec:example}
\begin{figure}[ht]
\begin{center}
\includegraphics[width=0.9\textwidth]{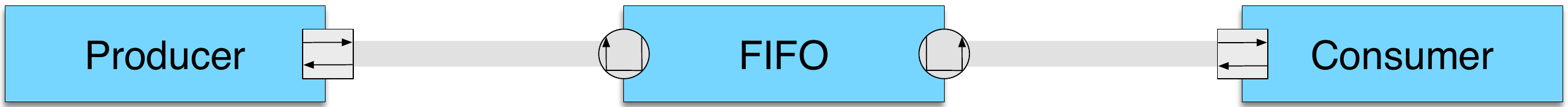}
\caption{Every 1 nanosecond, the producer writes 1 character
to the FIFO channel with probability $p_1$, while the consumer reads 1 character with
probability $p_2$. The FIFO channel is designed to ensure that all data is reliably delivered despite the
varying rates of production and consumption}
\label{fig:fifo}
\end{center}
\end{figure}

We will use a simple case study with a \emph{First-In First-Out} (FIFO) channel as a running example (see \figref{fig:fifo} with the graphical notations in~\cite{glm02}). This example illustrates how designers can create hierarchical channels that encapsulate both design structure and communication protocols.
In the design, once a nanosecond the producer will write one character to the FIFO channel with probability $p_1$, while the consumer will read one character from the FIFO channel with probability $p_2$.
The FIFO channel which is derived from \word{sc\_channel} encapsulates the communication protocol
between the producer and the consumer.

The FIFO channel is designed to ensure that all data is reliably delivered despite the
varying rates of production and consumption. The channel uses an event notification handshake protocol for both the input and output. It uses a circular buffer implemented within a static array to store and retrieve the items within the FIFO channel. We assume that the sizes of the messages and the FIFO buffer are fixed. Hence, it is obvious that the time required to transfer completely a message, or message \emph{latency}, depends on the production and consumption rates, the FIFO buffer size, the message size, and the probabilities of successful writing and reading. The full implementation of the example can be obtained at the website of our tool~\cite{pscv,nlq16}, in which the probabilities of writing and reading are implemented with the Bernoulli distributions with probabilities $p_1$ and $p_2$ respectively from \emph{GNU Scientific Library} (GSL)~\cite{gsl}.

The quantitative analysis under consideration is: What is the probability that each single message is transferred completely (e.g., including the message delimiters) within $\word{T_1}$ nanoseconds during $\word{T}$ nanoseconds of operation? This kind of analysis can, thus, be conducted in the early design steps. To formulate the underlying property more precisely, we have to take into account the agreement protocol between the producer and consumer, e.g., a simple protocol can be every message has special starting delimiter with the character '\&' and ending delimiter with the character '@'. Thus, the property can be translated in BLTL as follows:
\begin{displaymath}
\varphi = \word{G_{\leq T}}((\word{c\_read} = \; '\&') \rightarrow \word{F_{\leq T_1}}(\word{c\_read} = \; '@'))
\end{displaymath}
where $\word{c\_read}$ is the character read in the FIFO channel by the consumer. The input providing to the SMC checker is $\word{Pr}(\varphi)$. This property is expressed in terms of the characters read in the FIFO channel by the consumer, but the communication protocol between the producer and the consumer is abstracted at a very high level. It is an illustration of a type of property that can be checked on TLM specifications. Consider a concrete implementation of the communication protocol between the producer and the consumer at \emph{Register-Transfer Level} (RTL), if a property is satisfiable at the transaction level then it is so at the register-transfer level. Therefore, the verification of such a property at the transaction level can be connected to its counterpart at RTL in order to check the correctness of RTL implementations. 

\section{SMC for SystemC Models}
\label{sec:smcsystemc}
In order to apply SMC for SystemC models which exhibit timed and probabilistic characteristics, this section presents the concepts of state and execution trace for SystemC. It also shows that the operational semantics of this class of SystemC models can be considered as stochastic processes.
\subsection{SystemC Model State}
Temporal logic formulas are interpreted over execution traces and traditionally a trace has been defined as a sequence of states in the execution of a model. Thus before we can define an execution trace we need a precise definition of the state of a SystemC model simulation. We are inspired by the definition of system state in~\cite{tva10}, which consists of the state of the simulation kernel and the state of the SystemC model user-code. We consider the external libraries as black boxes, meaning that their states are not exposed.

Let $Sp$ be the set of simulation kernel phases, $Ev$ be the set of all events, and $X$ be the set of all variables in the model that includes all module attributes, ports, and channels. Let $F$ be the set of all functions in the model, a configuration of the call stack is represented by a tuple $(f, f_p, f_r)$ where $f \in F$ is the executing function, $f_p$ is its parameters, and $f_r$ is its return values. Let $Cs$ be the set of all call stack configurations. Let $Loc$ be the set of all locations, e.g., a specific statement reached during the execution and $Proc$ is the set of all statuses of all module processes in the model. Then a state is formally represented by a tuple $(sp, ev, \sigma, l, cs, proc)$ where:
\begin{itemize}
  \item $sp \in Sp$ is the current phase of the simulation kernel, e.g., delta-cycle notification, simulation-cycle simulation,
  \item $ev \in 2^{Ev}$ is the set of events that are notified during the execution of the model,
  \item $\sigma : X \rightarrow \mathbb{D}_X$  is a map from variables to its domain values,
  \item $l \in Loc$ is the current location,
  \item $cs \in Cs$ is the current configuration of the call stack,
  \item $proc \in Proc$ is the current status of the module processes, e.g., suspended or runnable.
\end{itemize}

We consider here some examples about states of the simulation kernel and the SystemC model user-code. Assume that a SystemC model has an event named $\word{e}$, then the model state can contain information such as the kernel is at the end of simulation-cycle notification phase and the event $\word{e}$ is notified. 

Consider the running example again, a state can be the information about the characters received by the consumer, represented by the variable with type $\type{char}$, $\word{c\_read}$. It also contains the information about the location of the program counter right before and after a call of the function $\word{send()}$ in the module $\word{Producer}$ that are represented by two Boolean variables $\word{send\_start}$ and $\word{send\_done}$, respectively. These variables hold the value $\word{true}$ immediately before and after a call of the function $\word{send()}$, respectively. 

Another example, we consider a module that has statements at different locations in the source code, in which these statements contain the division operator ``/'' followed by zero or more spaces and the variable ``$\word{a}$'' (e.g., the statement $\word{y = (x + 1) / a}$). Then, a Boolean variable which holds the value $\word{true}$ right before the execution of all such statements can be used as a part of the states.

Let $S$ be the (finite or infinite) set of all states of the model. We use $V = \{v_0,...,v_{n-1}\}$ to denote the finite set of variables of primitive type (e.g, usual scalar or enumerated type in C/C++) whose value domain $\mathbb{D}_V = \mathbb{D}_{v_0} \times \cdots \times \mathbb{D}_{v_{n-1}} \subseteq S$ represents the states of a SystemC model.

We have discussed so far the state of a SystemC model execution. It remains to discuss how the semantics of the temporal operators is interpreted over the states in the execution of the model. That means how the states are sampled in order to make the transition from one state to another state. The following definition gives the concept of \emph{temporal resolution}, in which the states are evaluated only at instances in which the temporal resolution holds. It allows the user to set granularity of time.
\begin{definition}[Temporal Resolution]
A temporal resolution $\temporal$ is a disjunction of a finite set of Boolean expressions defined over $V$. It specifies when the set of variables $V$ is evaluated to generate a new state.
\end{definition}
Temporal resolution can be used to define a more fine-grained model of time than a coarse-grained one provided by a cycle-based simulation. We call the expressions in $\temporal$ \emph{temporal events}. Whenever a temporal event is satisfied, or we say that the temporal event occurs, $V$ is sampled.

For example, in the producer and consumer model, assume that we want the satisfaction of the underlying BLTL formula $\varphi$ to be checked whenever at either the end of simulation-cycle notification or immediately after the event $\word{write\_event}$ is notified during a run of the model. Hence, we can define a temporal resolution as $\temporal = \{\word{end\_sc}, \word{we\_notified}\}$, where $\word{end\_sc}$ and $\word{we\_notified}$ are Boolean expressions that have the value $\word{true}$ whenever the kernel phase is at the end of the simulation-cycle notification and the event
$\word{write\_event}$ notified, respectively.

We denote the set of occurrences of temporal events from $\mathcal{T}_r$ along an execution of a SystemC model by $\mathcal{T}^{s}_r$, called a \textit{temporal resolution set}. The value of a variable $v \in V$ at an event occurrence $e_c \in \mathcal{T}^{s}_{r}$ is defined by a mapping $\xi^{v}_{val}: \mathcal{T}^{s}_{r} \rightarrow \mathbb{D}_{v}$, where $\mathbb{D}_{v}$ is the value domain of $v$.

A mapping $\xi_t: \mathcal{T}^{s}_{r} \rightarrow \mathcal{T}$ is called a \textit{time event} that identifies the simulation time at each occurrence of an event from the temporal resolution. Hence, the set of time points, called \textit{time tag}, which corresponds to a temporal resolution set $\mathcal{T}^{s}_{r} = \{e_{c_0},...,e_{c_{N-1}}\}, N \in \mathbb{N}$, is given as follows.
\begin{definition}[Time Tag]
Given a temporal resolution set $\mathcal{T}^{s}_{r}$, the \textit{time tag} $\mathcal{T}$ corresponding to $\mathcal{T}^{s}_{r}$ is a finite or infinite set of non-negative reals $\{t_0,t_1,...,t_{N-1}\}$, where $t_{i+1} - t_i = \delta t_i \in \mathbb{R}_{\geq 0}$ and $t_i = \xi_t(e_{c_i})$.
\end{definition}

Therefore, the state of the SystemC model at an event occurrence $e_c \in \mathcal{T}^{s}_{r}$ (with the corresponding simulation time $t = \xi_t(e_{c})$) is formally defined as follows.
\begin{definition}[Model State]
Let $V = \{v_0,...,v_{n-1}\}$ be a finite set of variables and $\temporal$ be a temporal resolution. Then the state of the model at the simulation time $t = \xi_t(e_{c})$, written $(s,t)$, is defined by a tuple $(\xi^{v_0}_{val},...,\xi^{v_{n-1}}_{val})$.
\end{definition}
\subsection{Model and Execution Trace}
A SystemC model can be viewed as a hierarchical network of parallel communicating processes. Hence, the execution of a SystemC model is an alternation of the control between the model's processes, the external libraries and the kernel process. The execution of the processes is supervised by the kernel process to concurrently update new values for the signals and variables with respect to the cycle-based simulation. For example, given a set of runnable processes in a simulation-cycle, the kernel chooses one of them to execute first as described in the prior section.

Let $V$ be the set of variables whose values represent the states of a SystemC model. The values of variables in $V$ are samples of the set of states $S$. A state can be considered as a random variable following an unknown probability distribution defined from all given probability distributions in the SystemC source code model. Given a temporal resolution $\mathcal{T}_r$ and its corresponding temporal resolution set along an execution of the model $\mathcal{T}^{s}_{r} = \{e_{c_0},...,e_{c_{N-1}}\}, N \in \mathbb{N}$, the evaluation of $V$ at the event occurrence $e_{c_i}$ is defined by the tuple $(\xi^{v_0}_{val},...,\xi^{v_{n-1}}_{val})$, or a state of the model at $e_{c_i}$, denoted by $V(e_{c_i}) = (V(e_{c_i})(v_0),V(e_{c_i})(v_1),...,V(e_{c_i})(v_{n-1}))$, where $V(e_{c_i})(v_k) = \xi^{v_k}_{val}(e_{c_i})$ with $k = 0,...,n-1$ is the value of the variable $v_k$ at $e_{c_i}$.

We denote the set of all possible evaluations by $V_{\mathcal{T}^{s}_{r}} \subseteq \mathbb{D}_V$, called the \textit{state space} of the random variables in $V$. State changes are observed only at the moments of event occurrences. Hence, the operational semantics of a SystemC model is represented by a \textit{stochastic process} $\{(V(e_{c_i}),\xi_t(e_{c_i})), e_{c_i} \in \mathcal{T}^{s}_{r}\}_{i \in \mathbb{N}}$, taking values in $V_{\mathcal{T}^{s}_{r}} \times \mathbb{R}_{\geq 0}$ and indexed by the parameter $e_{c_i}$, which are event occurrences in the temporal resolution set $\mathcal{T}^{s}_{r}$. An execution trace is a realization of the stochastic process and is given as follows.
\begin{definition}[Execution Trace]
An execution trace of a SystemC model corresponding to a temporal resolution set $\mathcal{T}^{s}_{r} = \{e_{c_0},...,e_{c_{N-1}}\}, N \in \mathbb{N}$ is a sequence of states and event occurrence times, denoted by $\omega = (s_0,t_0)...(s_{N-1},t_{N-1})$, such that for each $i \in 0,...,N-1$, $s_i = V(e_{c_i})$ and $t_i = \xi_t(e_{c_i})$.
\end{definition}
The value $N$ is called the length (finite or infinite) of the execution, also denoted by $|\omega|$. Given $V' \subseteq V$, the \emph{projection} of $\omega$ on $V'$, written $\omega \downarrow_{V'}$, is an execution trace such that $|\omega \downarrow_{V'}| = |\omega|$ and $\forall v \in V'$, $\forall e_c \in \mathcal{T}^{s}_{r}$, $V'(e_c)(v) = V(e_c)(v)$.
\subsection{Properties Expressing}
\label{sec:bltl}
Our framework accepts input properties of the forms $\word{Pr}(\varphi)$, $\word{Pr}_{\geq \theta}(\varphi)$, and $\word{X_{\leq T}}(rv)$, where $\varphi$ is a BLTL formula. The former is used to compute the probability that $\varphi$ is satisfied by the model. The latter asserts that this probability is at least equal to the threshold $\theta$. The last one returns the mean value of random variable $rv$.

The set of atomic propositions $\word{AP}$ in the logic which describes SystemC code features and the simulation semantics is a set of Boolean expressions defined over a subset of $V$ called \emph{observed variables} and the standard operators ($+,-,*,/,>,\geq,<,\leq,!=,=$). The semantics of the temporal operators in BLTL formulas interpreted over states is defined by a temporal resolution. In other words, the temporal resolution determines how the states are sampled in order to make the transition from one state to another state. The observed variables and temporal resolutions supported by the framework are summarized below; see the tool manual\footnote{ http://project.inria.fr/pscv/en/documentation/} for the full syntax and semantics. It should be noted that the current implementation only supports the use of simulation phases, locations, and events to define temporal resolutions. The implementation provides a mechanism that allows users to declare observed variables in order to define the set of propositions $\word{AP}$ via a high-level language in a configuration file as the input of our tool.
\begin{itemize}
\item \emph{Attribute}. Users can define an observed variable whose value and type are equal to the value and type of a module's attribute in the user code. Attributes can be public, protected, or private. For example, $\word{attribute} \; \word{a.t} \; \word{a\_t}$ defines a variable named \word{a\_t} whose value and type are equal to the value and type of the private attribute \word{t} of the module instance \word{a}.
\item \emph{Function}. Let \word{f()} be a C++ function with $k$ arguments in the user code. Users can refer to locations in the source code that contain the function call, immediately after the function call, immediately before the first executable statement, and immediately after the last executable statement in \word{f()} by using Boolean observed variables \word{f:call}, \word{f:return}, \word{f:entry}, and \word{f:exit}, respectively. Moreover, users can define an observed variable \word{f:i}, with $i = 0...k$, whose value and type are equal to the value and type of the return object (with $i = 0$) or $i^{th}$ argument of function \word{f()} before executing the first statement in the function body. For example, if the function $\word{int} \; \word{div(int \; x, \; int \; y)}$ is defined in the user code, then the formula $\word{G_{\leq T}} (\word{div:entry} \rightarrow \word{div:2} \; != 0)$ asserts that the divisor is nonzero whenever the $\word{div}$ function starts execution.
\item \emph{Simulation Phase}. There are 18 predefined Boolean observed variables which refer to the 18 kernel states~\cite{pscv}. These variables are usually used to define a temporal resolution. For example, the formula $\word{G_{\leq T}} (p = 0)$ which is accompanied with the temporal resolution $\word{MON\_DELTA\_CYCLE\_END}$ requires the value of variable $p$ to be zero at the end of every delta-cycle.
\item \emph{Event}. For each SystemC event $e$, the framework provides a Boolean observed variable $\word{e.notified}$ that is true only when the simulation kernel actually notifies $e$. For example, the formula $\word{G_{\leq T}} (\word{e.notified})$ which is accompanied with the temporal resolution $\word{MON\_UPDATE\_PHASE\_END}$ says that the event $e$ is notified at the end of every update phase.
\end{itemize}
Referring to the running example, the declaration \word{location} \word{send\_start} \word{"\%Producer} \word{::}\word{send()"}\word{:call} declares a Boolean variable \word{send\_start} that holds the value \word{true} immediately before the execution of the model reaches a call site of the function \word{send()} in the module \word{Producer}. The characters received by the consumer which is represented by the variable \word{c\_read} can be declared as \word{attribute} \word{pnt\_con} $\rightarrow$ \word{c\_int} \word{c\_read}, where \word{pnt\_con} is a pointer to the \word{Consumer} object and \word{c\_int} is an attribute of the \word{Consumer} module representing the received character. Then the considered property of the running example can be constructed by using the following predicates defined over the variable \word{c\_read}: $\word{c\_read} = '\&'$ and $\word{c\_read} = '@'$.

Another example, assume that we want to answer the following question: \emph{``Over a period of $\word{T}$ time units, is the probability that the number of elements in the FIFO buffer is between $n_1$ and $n_2$ greater or equal to $\theta$ with the confidence $1 - \alpha$''}? The predicates that need to be defined in order to construct the underlying BLTL formula are $n_1 \leq \word{n_{elements}}$ and $\word{n_{elements}} \leq n_2$, where $\word{n_{elements}}$ is an integer variable that represents the current number of elements in the FIFO buffer (it captures the value of the \word{num\_elements} attribute in the \word{Fifo} module). Then, the property is translated in BLTL with the operator ``always'' as follows. $\word{Pr}_{\geq \theta}(\varphi)$ and the confidence $1 - \alpha$ are given as inputs to the checker.
\begin{displaymath}
\varphi = \word{G_{\leq T}}((n_1 \leq \word{n_{elements}}) \; \& \; (\word{n_{elements}} \leq n_2))
\end{displaymath}

\section{Implementation}
\label{sec:implementation}
\begin{figure}[ht]
\begin{center}
\includegraphics[width=0.98\textwidth]{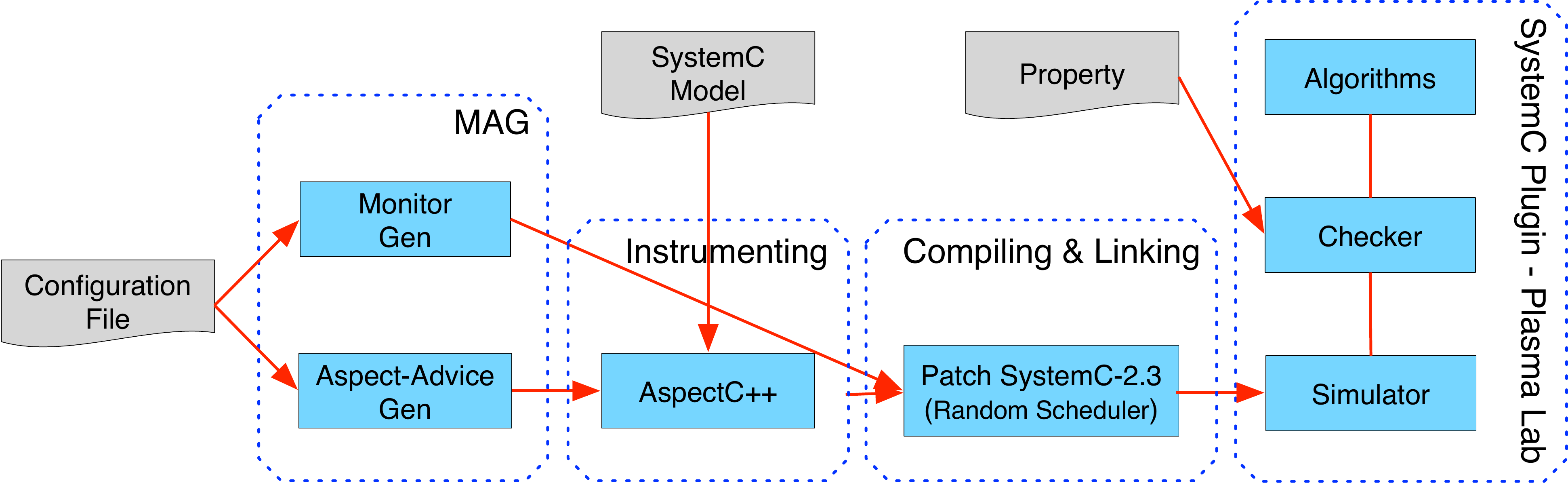}
\caption{PSCV consists of off-the-self components (AspectC++ and Plasma Lab),
modified component (patch SystemC-2.3.0), and original components (MAG and SystemC plugin)}
\label{fig:architecture}
\end{center}
\end{figure}

We have implemented an SMC-based verification tool~\cite{NgoLJ16,nlq16}, PSCV, that contains two main original components: a \emph{Monitor} and \emph{Aspect-advice Generator} (MAG) and a \emph{Statistical Model Checker} (SystemC Plugin). The tool whose architecture is depicted in \figref{fig:architecture} is considered as a runtime verification tool for probabilistic temporal properties.
\subsection{The Architecture}
It consists of the following off-the-self, modified and original components.
\begin{itemize}
\item An off-the-self component, AspectC++~\cite{gss01}, a C++ \emph{Aspect Oriented Programming} (AOP) compiler for instrumenting the MUV.
\item A modified component, a patch SystemC-2.3.0 for facilitating the communication between the kernel and the monitor and implementing a random scheduler.
\item Two original components are MAG, a C++ tool for automatically generating monitor and aspect-advices for instrumentation, and SystemC plugin, a plugin of the statistical model checker Plasma Lab~\cite{bcl13}.
\end{itemize}
\subsubsection{Execution Trace Extraction}
In principle, the full state can be observed during the simulation of the model. In practice, however, users define a set of variables of interest, according to the properties that the users want to verify, and only these variables appear in the states of an execution trace.
Given a SystemC model, we use $V_{\word{obs}} \subseteq V$ to denote the set of variables, called \emph{observed variables}, to expose the states of the SystemC model. Then, the observed execution traces of the model are the projections of the execution traces on $V_{\word{obs}}$, meaning that for every execution trace $\omega$, the corresponding observed execution trace is $\omega \downarrow_{V_{\word{obs}}}$. In the following, when we mention execution traces, we mean observed execution traces.

In PSCV, based on the techniques in~\cite{tva10}, the set of observed variables and temporal resolution are converted into a C++ monitor class and a set of aspect-advices. MAG generates three files: \word{aspect\_definitions.ah} containing a set of AspectC++ \emph{aspect} definitions, \word{monitor.h}, and \word{monitor.cc} implementing a monitor as a C++ class and another class called \word{local\_observer} that is responsible for invoking the callback functions. The callback functions will invoke the sampling function at the right time point during the MUV simulation.

The monitor has \word{step()}, a sampling function. It waits for a request from the SystemC Plugin. If the request is stopping the current simulation, it then terminates the MUV's execution. If the plugin requests a new state, then the current values of all observed variables and the simulation time are sent. The \word{step()} function is called at every time point defined by the temporal resolution. These time points can be kernel phases, event notifications, or locations in the MUV code control flow. In such cases, the patch SystemC kernel needs to communicate with the \word{local\_observer}, e.g., when a delta-cycle ends, via a class called \word{mon\_observer} to invoke the \word{step()} function of the monitor. In case of locations in the user-code, the advice code generated by MAG will call the callback function to invoke the \word{step()} function.

\begin{figure}[!th]
\includegraphics[width=1.0\textwidth]{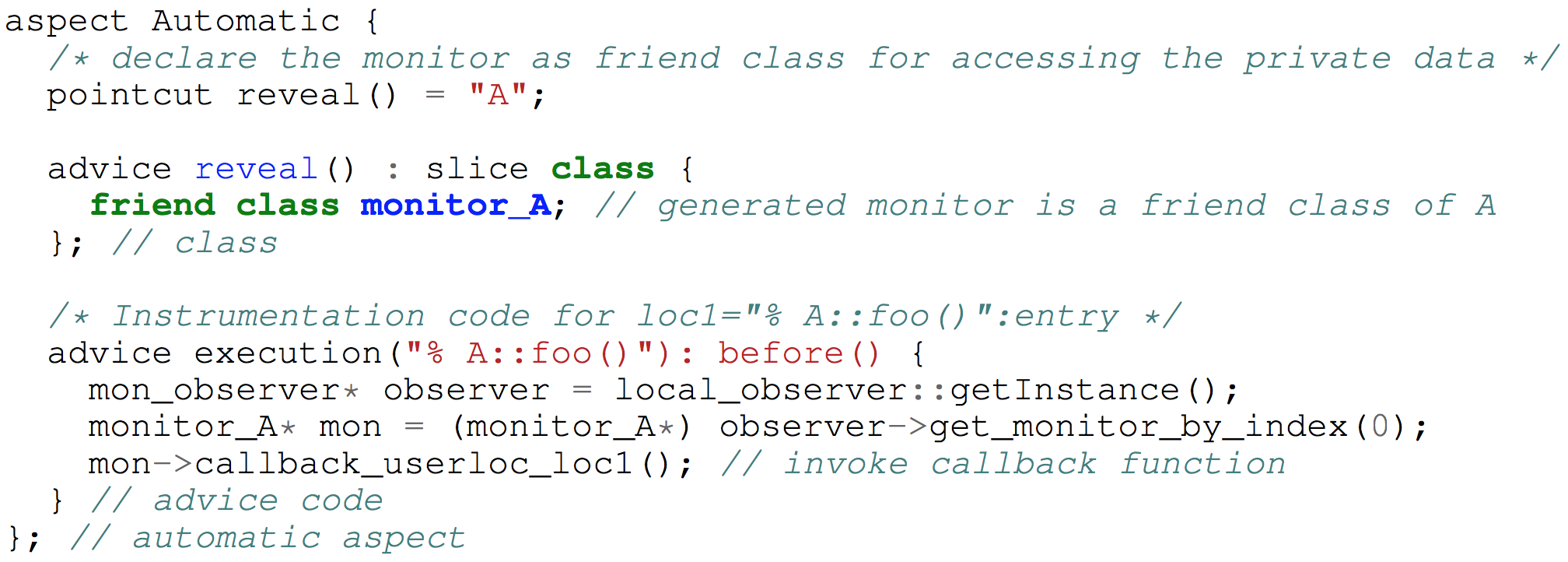}
\caption{Example of generated AspectC++ aspect.}
\label{lst:aspect}
\end{figure}

The aspect in AspectC++ is an extension of the class concept of C++ to collect advice code implementing a common crosscutting concern in a modular way. For example, to access all attributes of a module called $\word{A}$ and the location immediately before the first executable statement of the function $\word{foo}$ in $\word{A}$ (defined in the configuration file with the syntax \word{\% \; A::foo():entry}). MAG will generate the aspect definition in \figref{lst:aspect}. The full syntax and semantics of aspects can be found at the web-site of AspectC++~\cite{acdoc}.
\subsubsection{Statistical Model Checker}
The statistical model checker is implemented as a plugin of Plasma Lab~\cite{bcl13} that establishes a communication, in which the generated monitor transmits execution traces of the MUV. In the current version, the communication is done via the standard input and output. When a new state is requested, the monitor reports the current state (the values of observed variables and the current simulation time) to the plugin. The length of traces depends on the satisfaction of the formula to be verified, which is finite due to the bounded temporal operators.

Similarly, the required number of traces depends on the statistical algorithms in use (e.g., sequential hypothesis testing or 2-sided Chernoff bound). The full set of algorithms implemented in Plasma Lab can be found at~\cite{plasma}.
\subsubsection{Random Scheduler}
\begin{figure}[!th]
\includegraphics[width=0.80\textwidth]{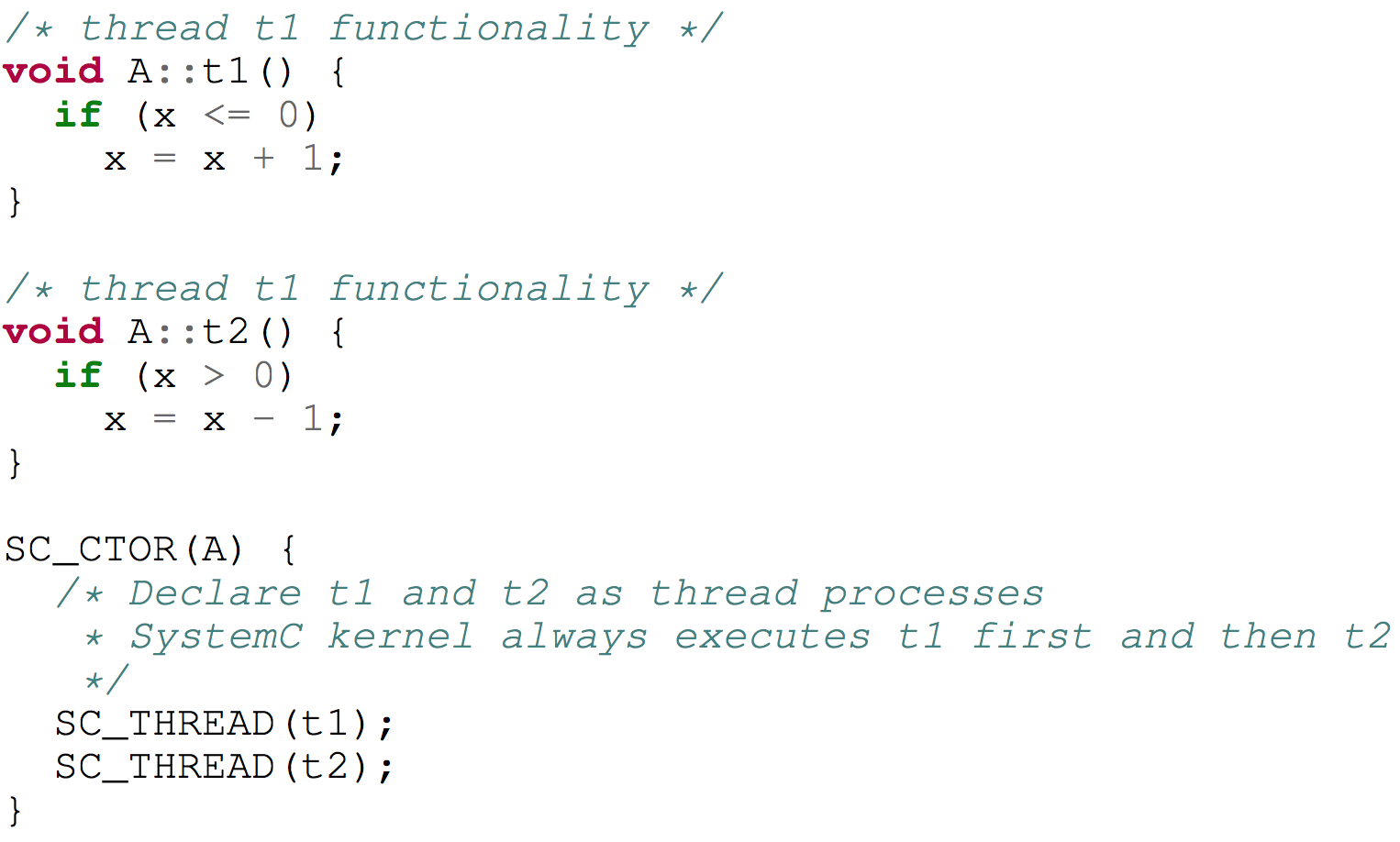}
\caption{Example of deterministic execution order.}
\label{lst:random}
\end{figure}

Verification does not only depend on the probabilistic characteristics of the MUV, but it also can be significantly affected by \emph{scheduling policy}. Consider a simple module $\word{A}$ consisting of two thread processes as shown in \figref{lst:random}, where $x$ is initialized to be $1$.
Assume that we want to compute the probability that $x$ is always equal to $1$. Obviously, $x$ depends on the execution order of two threads, e.g., its value is $1$ if $\word{t2}$ is executed before the execution of $\word{t1}$ and $0$ if the order is $\word{t1}$ then $\word{t2}$.

The current scheduling policy is deterministic, in which it always picks the process that is first added into the queue (the current open-source SystemC kernel implementation uses a queue to store a set of runnable processes). Hence, only one execution order, $\word{t1}$ then $\word{t2}$, is verified instead of two possible orders. As a result, the computed probability is $0$, however, ideally it should be $0.5$. Therefore, it is more interesting if a verification is performed on all possible execution orders than a fixed one. In many cases, there is no decision or \`a priori knowledge of the scheduling to be implemented. Moreover, verification of a specification should be independent of the scheduling policy to be finally implemented.

To make our verification perform on all possible execution orders of the MUV, we have implemented a random scheduler by patching the current open-source SystemC kernel implementation. The source of process scheduling is the evaluation phase, in which one of the runnable processes will be executed. Given a set of $N$ runnable processes in the queue every time at the evaluation phase, our random scheduler randomly chooses one of them to execute. The random algorithm is implemented by generating a random integer number uniformly over a range $[0,N-1]$. For more simulation efficiency and implementation simplicity, we employ the \word{rand()} function and $\%$ operator in C/C++.

\subsection{The Verification Flow}
\begin{figure}[ht]
	\begin{center}
		\includegraphics[width=0.95\textwidth]{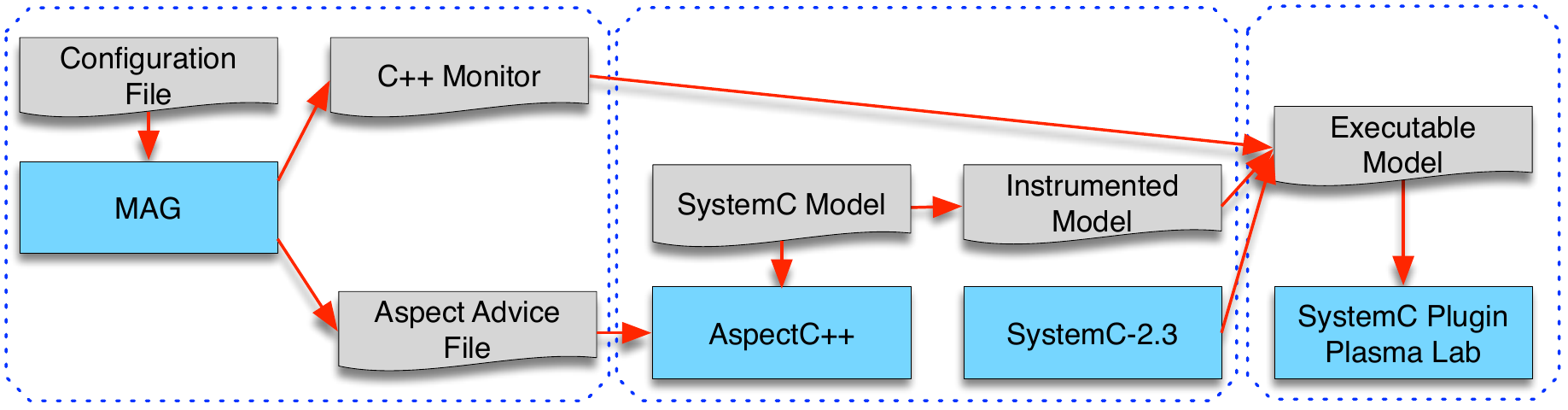}
		\caption{The verification flow consists of 3 steps: writing a configuration file,
		instrumenting the MUV using AspectC++, and verifying the instrumented MUV using the SystemC plugin}
		\label{fig:flow}
	\end{center}
\end{figure}

The verification flow using PSCV consists of three steps, as shown in \figref{fig:flow}.
In the first step, users write a configuration file containing a set of typed variables (\emph{observed variables}), a Boolean expression (\emph{temporal resolution}), and all properties to be verified. MAG translates the configuration file into a C++ monitor and a set of aspect-advices. The set of aspect-advices is then used by AspectC++ as input to automatically instrument the MUV to expose the user-code’s state and simulation kernel's state in the second step. The instrumented model and the generated monitor are compiled and linked together with the modified SystemC kernel into an executable model.

Referring to the running example, users can define the set of observed variables $V_{\word{obs}} = \{\word{c\_read}, \word{n_{elements}}\}$, where \word{c\_read} is the character read in the FIFO buffer and $\word{n_{elements}}$ is the number of characters in the FIFO buffer. The temporal resolution will be defined as $\temporal = \{\word{MON\_TIMED\_NOTIFY\_PHASE\_END}\}$, where \word{MON\_TIMED\_NOTIFY\_PHASE\_END} is one of $18$ predefined Boolean observed variables which refer to the 18 kernel states~\cite{pscv}. That means that a new state in execution traces is produced whenever the simulation kernel is at the end of simulation-cycle notification phase or every one nanosecond in the example since the time resolution is one nanosecond. The full configuration file is included in the source code of the example\footnote{http://project.inria.fr/pscv/producer-and-consumer/}.

Finally, SystemC plugin, a plugin of the statistical model checker Plasma Lab~\cite{bcl13}, simulates independently the executable model in order to make the monitor produce execution traces with inputs provided by users. The inputs can be generated using any standard stimuli-generation technique. These traces are finite in length since the BLTL semantics~\cite{sva05} is defined with respect to finite execution traces. The number of simulations is determined by the statistical algorithm used by the plugin based on the absolute error and the level of confidence. Given these execution traces and the user-defined absolute error and confidence, the SystemC plugin employs the statistical model checking to produce an estimation of the probability that the property is satisfied or an assertion that this probability is at least equal to a threshold.

\section{Experimental Results}
\label{sec:casestudy}
We report the experimental results for the running example and also demonstrate the use of our verification tool to analyze the dependability of a large embedded control system. The number of components in this system makes numerical approaches such as PMC unfeasible. In both case studies, we used the 2-sided Chernoff bound algorithm with the absolute error $\epsilon = 0.02$ and the confidence $1 - \alpha = 0.98$. The experiments were run on a machine with Intel Core i7 2.67 GHz processor and 4GB RAM under the Linux OS with SystemC 2.3.0, in which the checking of the properties in the running example took from less than one minute to several minutes. The analysis of the embedded and control system case study took almost $2$ hours, in which $90$ properties were verified.
\subsection{Producer and Consumer}
Let us go back to the running example in Section \ref{sec:example}, recall that we want to compute the probability that the following property $\varphi$ satisfies every $1$ nanosecond, with the absolute error $0.02$ and the level of confidence $0.98$. In this verification, both the FIFO buffer size and message size are $10$ characters including the starting and ending delimiters, and the production and consumption rates are $1$ nanosecond.
\begin{displaymath}
\varphi = \word{G_{\leq T}}((\word{c\_read} = \; '\&') \rightarrow \word{F_{\leq T_1}}(\word{c\_read} = \; '@'))
\end{displaymath}

First, we compute $\word{Pr}(\varphi)$ with various values of $p_1$ and $p_2$. The results are given in Table~\ref{tab:fifolatency} with $\word{T} = 5000$ and $\word{T_1} = 25$ nanoseconds. Obviously, the probability that the message latency is smaller than $\word{T_1}$ time units increases when $p_1$ and $p_2$ increase, where $p_1$ and $p_2$ are probabilities that the producer write to and consumer reads from the FIFO channel successfully. That means in general the latency is shorter when either the probability that the producer successfully writes to the FIFO channel increases, or the probability that the consumer successfully reads increases.
\begin{table}[!ht]
\centering
\begin{tabular}{c|ccc}
$p_1$\textbackslash $p_2$ & $0.3$ & $0.6$ & $0.9$\\
\hline
$0.6$ & $0$ & $0.0194$ & $0.0720$\\\\
$0.9$ & $0$ & $0.0835$ & $1$\\
\end{tabular}
\caption{The probability that the message latency is smaller than $25$ in the first $5000$ nanoseconds of operation}
\label{tab:fifolatency}
\end{table}

\begin{figure}[!ht]
\begin{center}
\includegraphics[width=0.95\textwidth]{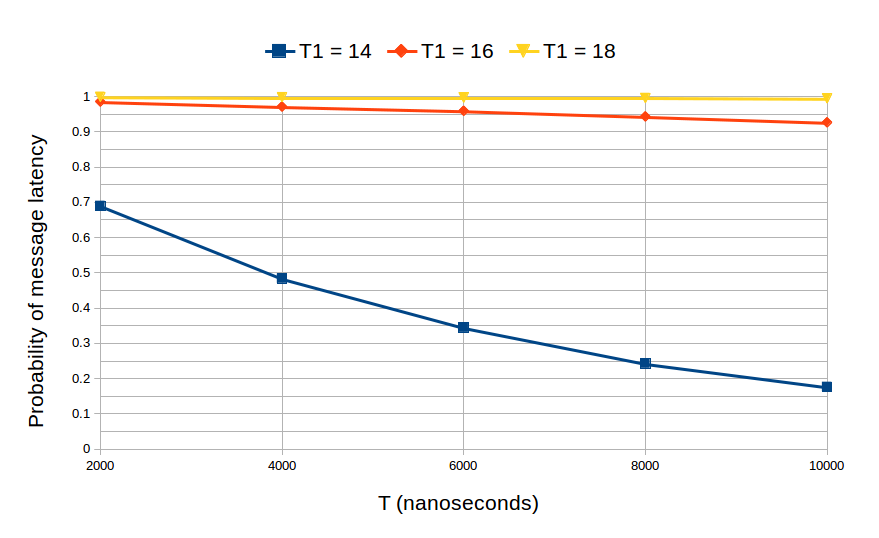}
\caption{The probability that the message latency is smaller than $\word{T_1}$ in the first $\word{T}$ nanoseconds of operation}
\label{fig:fifolatency}
\end{center}
\end{figure}

Second, we compute the probability that a message is sent completely (or the message latency) from the producer to the consumer within $\word{T_1}$ time units over a period of $\word{T}$ time units of operation, in which the probabilities $p_1$ and $p_2$ are fixed at $0.9$. Fig.~\ref{fig:fifolatency} shows this probability with different values of $\word{T_1}$ over $\word{T} = 10000$ nanoseconds. It is observed that the message latency is just below $18$ nanoseconds.
\subsection{Embedded Control System}
\label{subsec:ecs}
This case study is closely based on the one presented in~\cite{mct94,knp07} but contains many more components. The system consists of an input processor ($\word{I}$) connected to $50$ groups of $3$ sensors, an output processor ($\word{O}$), connected to $30$ groups of $2$ actuators, and a main processor ($\word{M}$), that communicates with $\word{I}$ and $\word{O}$ through a bus. At every cycle of $1$ minute, the main processor polls data from the input processor that reads and processes data from the sensor groups. Based on this data, the main processor constructs commands to be passed to the output processor for controlling the actuator groups.

The reliability of the system is affected by the failures of the sensors, actuators, and processors. The probability of bus failure is negligible, hence we do not consider it. The sensors and actuators are used in $37-\text{of}-50$ and $27-\text{of}-30$ modular redundancies, respectively. That means if at least $37$ sensor groups are functional (a sensor group is functional if at least $2$ of the $3$ sensors are functional), the system obtains enough information to function properly. Otherwise, the main processor is reported to shut the system down. In the same way, the system requires at least $27$ functional actuator groups to function properly (an actuator group is functional if at least $1$ of the $2$ actuators is functional). Transient and permanent faults can occur in processors $\word{I}$ or $\word{O}$ and prevent the main processor($\word{M}$) to read data from $\word{I}$ or send commands to $\word{O}$. In that case, $\word{M}$ skips the current cycle. If the number of continuously skipped cycles exceeds the limit $\word{K}$, the processor $\word{M}$ shuts the system down. When a transient fault occurs in a processor, rebooting the processor repairs the fault. Lastly, if the main processor fails, the system is automatically shut down. The mean times to failure for the sensors, the actuators, and the processors are 1 month, 2 months and 1 year, respectively. The mean time to transient failure is 1 day and I/O processors take 30 seconds to reboot. In the model we represent 30 seconds as 1 time unit.

The reliability of the system is modeled as a CTMC~\cite{mge82,tks82,gaa87} that is realized in SystemC, in which a sensor group has $4$ states ($0, 1, 2, 3$, the number of working sensors), $3$ states ($0, 1, 2$, the number of working actuators) for an actuator group, $2$ states for the main processor ($0$: failure, $1$: functional), and $3$ states for I/O processors ($0$: failure, $1$: transient failure, $2$: functional). A state of the CTMC is represented as a tuple of the component's states, and the mean times to failure define the delay before which a transition between states is enabled. The delay is sampled from a negative exponential distribution with parameter equal to the corresponding mean time to failure. Hence, the model has about $2^{155}$ states compared to the model in~\cite{knp07} with about $2^{10}$ states, that makes the PMC technique unfeasible. That means the state explosion likely occurs, even if some abstraction, e.g., symbolic model checking is applied. The full implementation of the SystemC code and experiments of this case study can be obtained at the website of our tool\footnote{\url{https://project.inria.fr/pscv/embedded-control-system/}}.

We define four types of failures: $\word{failure_1}$ is the failure of the sensors, $\word{failure_2}$ is the failure of the actuators, $\word{failure_3}$ is the failure of the I/O processors and $\word{failure_4}$ is the failure of the main processor. For example, $\word{failure_1}$ is defined by $(\word{number\_sensors} < 37) \wedge (\word{proci\_status} = 2)$. It specifies that the number of working sensor groups has decreased below $37$ and the input processor is functional, so that it can report the failure to the main processor. \word{number\_sensors} and \word{proci\_status} are observed variables that have same type and value as the number of working sensor groups and the current status of the input processor, respectively. We define $\word{failure_2}$, $\word{failure_3}$, and $\word{failure_4}$ in a similar way.

\begin{figure}[!ht]
\begin{center}
\includegraphics[width=0.95\textwidth]{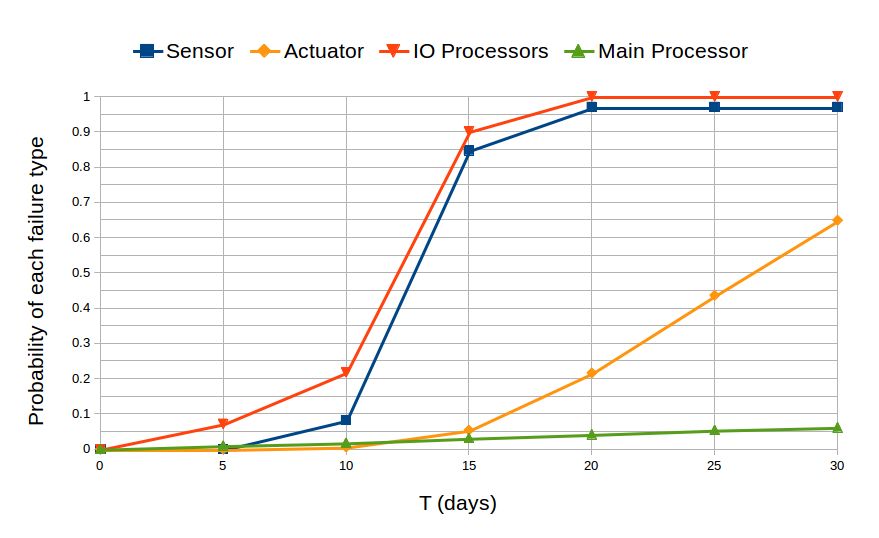}
\caption{The probability that each of the 4 failure types is the cause of system shutdown in the first $\word{T}$ time of operation}
\label{fig:failure}
\end{center}
\end{figure}

Our analysis is based on the one in~\cite{knp07} with $\word{K} = 4$ where the properties are checked every $1$ time unit. First, we study the probability that each of the four types of failure eventually occurs in the first $\word{T}$ time units of operation. This is done using the following BLTL formula $\word{F_{\leq T}} (\word{failure_i})$. Fig.~\ref{fig:failure} plots these probabilities over the first $30$ days of operation. We observe that the probabilities that the sensors and I/O processors eventually fail are more than the probability that the other components do. In the long run, they are almost the same and approximate to $1$, meaning that the sensors and I/O processors will eventually fail with probability $1$. The main processor has the smallest probability to eventually fail.
\begin{figure}[!ht]
\begin{center}
\includegraphics[width=0.95\textwidth]{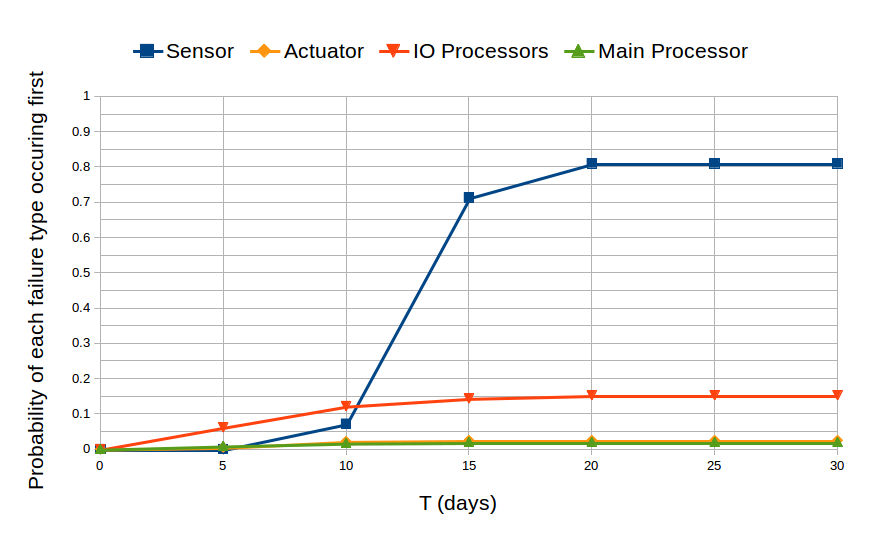}
\caption{The probability that each of the 4 failure types is the cause of system shutdown in the first $\word{T}$ time of operation}
\label{fig:failurefirst}
\end{center}
\end{figure}

Second, we try to determine which kind of component is more likely to cause the failure of the system, meaning that we determine the probability that a failure related to a given component occurs before any other failures. The atomic proposition $\word{shutdown} = \bigvee_{i=1}^4\word{failure_i}$ indicates that the system has shut down because one of the failures has occurred, and the BLTL formula $\neg \word{shutdown} \word{U_{\leq T}} \word{failure_i}$ states that the failure $i$ occurs within $\word{T}$ time units and no other failures have occurred before it. Fig.~\ref{fig:failurefirst} shows the probability that each kind of failure occurs first over a period of $30$ days of operation. It is obvious that the sensors are likelier to cause a system shutdown. At $\word{T} = 20$ days, it seems that we reached a stationary distribution indicating for each kind of component the probability that it is responsible for the failure of the system.

\begin{figure}[!ht]
\begin{center}
\includegraphics[width=0.95\textwidth]{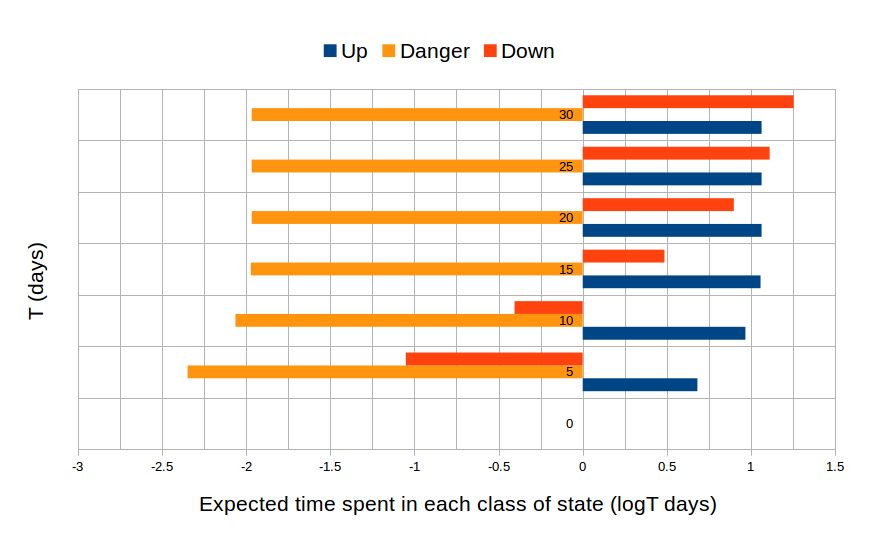}
\caption{The expected amount of time spent in each of the states: ``up'', ``danger'' and ``shutdown''}
\label{fig:timespent}
\end{center}
\end{figure}

For the third part of our analysis, we divide the states of system into three classes: ``up'', where every component is functional, ``danger'', where a failure has occurred but the system has not yet shut down (e.g., the I/O processors have just had a transient failure but they have rebooted in time), and ``shutdown'', where the system has shut down~\cite{knp07}. We aim to compute the expected time spent in each class of states by the system over a period of $\word{T}$ time units. To this end, we add in the model, for each class of state $\word{c}$, a random variable $\word{reward\_c}$ that counts the number of times the system is in the state $\word{c}$. Hence, it measures the time spent in the class $\word{c}$. 
In our tool, the formula $\word{X_{\leq T}} \word{reward\_c}$ returns the mean value of $\word{reward\_c}$ after $\word{T}$ time of execution. The results are plotted in Fig.~\ref{fig:timespent}. From $\word{T} = 20$ days, it seems that the amounts of time spent in the ``up'' and ``danger'' states are converged at $10^{1.063} = 11.57$ days and $10^{-1.967} = 0.01$ days, respectively. 

\begin{figure}[ht]
\centering
\includegraphics[width=0.95\textwidth]{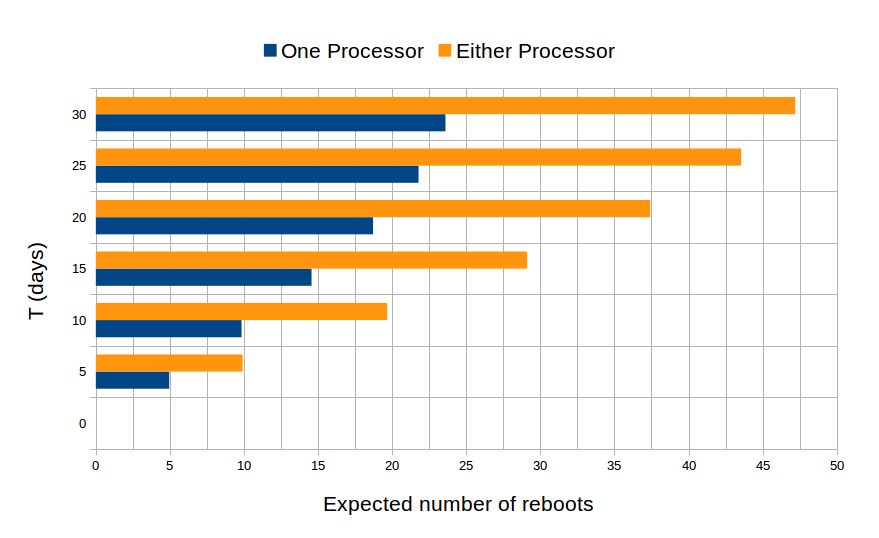}
\caption{Expected number of reboots that occur in the first $\word{T}$ time of operation}
\label{fig:numberofreboots}
\end{figure}

\begin{figure}[ht]
\centering
\includegraphics[width=0.95\textwidth]{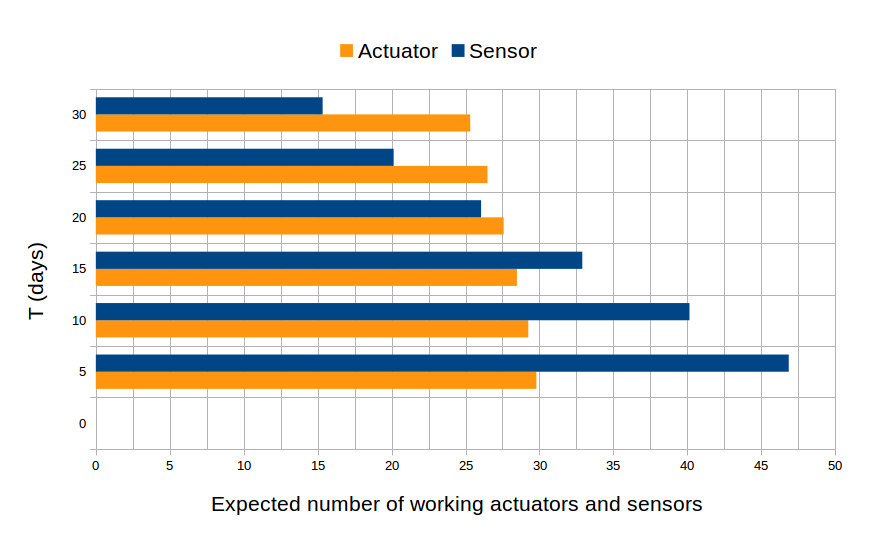}
\caption{Expected number of functional sensor and actuator groups in the first $\word{T}$ time of operation}
\label{fig:numberofworkinggroups}
\end{figure}

Finally, we approximate the number of reboots of the I/O processors, and the number of functional sensor groups and actuator groups over time by computing the expected values of random variables that count the number of reboots, functional sensor and actuator groups. The results are plotted in Fig. \ref{fig:numberofreboots} and Fig. \ref{fig:numberofworkinggroups}. It is obvious that the number of reboots of both processors is double the number of reboots of each processor since they have the same behavior model.
\subsection{Random Scheduler}
\label{subsec:scheduler}
This case study\footnote{\url{https://project.inria.fr/pscv/random-scheduler-examples/}} consisting of $3$ examples evaluates our random scheduler implementation. In example $1$ and example $2$, we implement a simple module with $2$ thread processes. The processes in the example $1$ have no synchronization point using \word{wait} statement, while there is one synchronization point in the example $2$. When these examples are run, we can see that there are $2$ execution orders in the first example and there are in total $4$ execution orders in the second example.

Example $3$ implements a SystemC model containing $3$ threads. Each thread has $2$ synchronization points using \word{wait} statements with other $2$ threads. As a result, each thread has $3$ execution segments. Therefore, there are totally $(3!)^3 = 216$ execution orders. In the first evaluation, we run the model $216$ times with our random scheduler in order to compute its \emph{coverage}. We got about $136$ different execution orders or the scheduler coverage is $136/216 \sim 63\%$. Second, we run the model until we get all $216$ different execution orders. Then on average we need to run the model $1382$ times. In other words, the \emph{dynamic random verification efficiency} of the random scheduler is $216/1382 \sim 15.6\%$. It seems that the implementation with the pseudo random number generator (PRNG), by using the \word{rand()} function, and $\%$ operator in C/C++ is not efficient. We are planning to investigate the Mersenne Twister generator~\cite{mn98} that is by far the most widely used general-purpose PRNG in order to better deal with this issue.

\section{Related Work}

Some work has been carried out for analyzing stochastic systems with PMC, for example, the dependability analysis of control system with PRISM~\cite{knp07}. PRISM supports construction and analysis of models as Markov chains, determining whether the model satisfies each property expressed in LTL. For example, the exact probabilities can be computed by PRISM. However, the main drawback of this approach is that when it deals with real-world large systems which make the PMC technique is unfeasible, even with some abstraction, i.e., symbolic model checking, is applied.

SMC has been recently applied in a wide range of research areas including software engineering (e.g., verification of critical embedded systems)~\cite{hwz08,sva051,ysi06}, system biology~\cite{jcj09}, or medical area~\cite{cmacs}. For instance, in~\cite{cdl08}, a framework of verifying properties of mixed-signal circuits, in which analog and digital quantities interact, was studied. The bounded linear temporal logic is used to express the properties, then the design of circuit and the verifying properties is connected to a statistical model checker. This approach consists of evaluating the properties on a representative subset of behaviors, generated by simulation, and answering the question whether the circuit satisfies the properties with a probability greater than or equal to a threshold.

Several verification techniques have been proposed for SystemC, for example, SystemC Verification Working Group proposed the \emph{SystemC Verification Standard} (SCV)~\cite{sysc} that provides APIs to verify functionality of programs. However, SCV does not mention temporal specifications. One can use SCV complementary to our verification framework in terms of automatically generating inputs for the MUV.

One of the earliest attempts at checking temporal specification properties for SystemC models is carried out by Braun et al.~\cite{bgr02}. The temporal properties are expressed as \emph{Finite Linear Temporal Logic} (FLTL) and the temporal resolution is defined by the boundary of the simulation-cycle notification. FLTL is linear temporal logic that interprets formulas over finite traces and supports temporal operators with respect to the temporal resolution. Then they proposed two approaches to extend test-bench features to SystemC for checking temporal properties: the checking is implemented directly within the SystemC language as an add-on library as SystemC itself. The other approach is to interface SystemC with an existing external test-bench environment, TestBuilder~\cite{cad01}. As described in~\cite{tva12}, the temporal resolution defined at boundary of the simulation-cycle notification is inadequate for SystemC verification, since it does not expose fully the semantics of the SystemC simulator kernel, which gives much finer grained temporal resolution. For example, the simulation may consist of a single delta-cycle if it is driven by immediate event notifications. As a result, the simulation clock never advances.

There has been a lot of work on the formalization of SystemC~\cite{hfg08,mmh08,Zhu2015, Zeng:2014,Zeng:2013}. In general, the goal of the formalization process is to extract a formal model from a SystemC program, so that tools like model-checkers can be applied. For example, Zeng et al. in~\cite{Zeng:2013} translate a subset of the operational semantics of SystemC as a guarded assignment system and use this translated model as an input for symbolic executors and checkers. However, all these formalizations consider semantics of SystemC and its simulator in some form of \textit{global model}, and they also suffer from the state space explosion when dealing with industrial and large systems.

Tabakov et al.~\cite{tva10,tva12} proposed a dynamic-analysis-based framework for monitoring temporal properties of SystemC models. This framework allows users to express the verifying properties by fully exposing the semantics of the simulator as well as the user-code. They extend LTL by providing some extra primitives for stating the atomic propositions and let users define a much finer temporal resolution. Their implementation consists of a modified simulation kernel, and a tool to automatically generate the \emph{monitors} and aspect-advices for instrumenting SystemC programs automatically with AOP.

\section{Conclusion}
\label{sec:conclusion}

This article presents the first attempt to verify non-trivial probabilistic and temporal properties of SystemC model with statistical model checking techniques. In comparison with the probabilistic technique, our technique allows us to handle large industrial systems modeling in SystemC as well as to expose a rich set of user-code primitives by automatically instrumenting the user-code with AspectC++. The framework contains two main components: a \emph{generator} that automatically generates a monitor and instruments the MUV based on the properties to be verified, and a \emph{statistical model checker} implementing a set of hypothesis testing algorithms. In comparison to the probabilistic model checking, our approach allows users to handle large industrial systems, expose a rich set of user-code primitives in the form of atomic propositions in BLTL, and work directly with SystemC models. For instance, our verification framework is used to analyze the dependability of large industrial computer-based control systems as shown in the case study.

Currently, we consider an external library as a ``black box'', meaning that we do not consider the states of external libraries. Thus, arguments passed to a function in an external library cannot be monitored. For future work, we would like to allow users to monitor the states of the external libraries. We also plan to apply statistical model checking to verify temporal properties of SystemC-AMS (Analog/Mixed-Signal).

To improve the dynamic random verification efficiency of the random scheduler's implementation, we are also planning to use the Mersenne Twister generator~\cite{mn98} that is by far the most widely used general-purpose PRNG instead of the default C/C++ generator.

\bibliographystyle{abbrv}
\bibliography{main}

\end{document}